# Predicting Indian stock market using the psycho-linguistic features of financial news

B. Shravan Kumar[1,2], Vadlamani Ravi[1, *] and Rishabh Miglani[3]

[1] Center of Excellence in Analytics,
Institute for Development and Research in Banking Technology (IDRBT),
Castle Hills Road #1, Masab Tank, Hyderabad-500057, India.

[2] School of Computer and Information Sciences, University of Hyderabad,
Hyderabad-500046, India.
`bangaaris@gmail.com,padmarav@gmail.com`

[3] Department of Mathematics, Indian Institute of Technology Kharagpur,
Kharagpur-721302, India.
`rishabh.miglani56@gmail.com`

**Abstract**

Financial forecasting using news articles is an emerging field. In this paper, we proposed hybrid intelligent models for stock market prediction using the psycholinguistic variables (LIWC and TAALES) extracted from news articles as predictor variables. For prediction purpose, we employed various intelligent techniques such as Multilayer Perceptron (MLP), Group Method of Data Handling (GMDH), General Regression Neural Network (GRNN), Random Forest (RF), Quantile Regression Random Forest (QRRF), Classification and regression tree (CART) and Support Vector Regression (SVR). We experimented on the data of 12companies' stocks, which are listed in Bombay Stock Exchange (BSE). We employed chi-squared and maximum relevance and minimum redundancy (MRMR) feature selection techniques on the psycho-linguistic features obtained from the new articles etc. After extensive experimentation, using Diebold-Mariano test, we conclude that GMDH and GRNN are statistically the best techniques in that order with respect to the MAPE and NRMSE values

.*Keywords*: Text mining; Stock market prediction; News articles; Psycholinguistic features; MAPE; NRMSE

## 1. Introduction

Stock Market prediction is an interesting research problem where stock value always varies significantly with respect to time. The time series is noisy and chaotic. Any forecasting model that finds the intricate relationship between the financial news about a company and its stock price is useful. Future stock values are predicted using the financial news about that company.

Especially, the outcome of the prediction (Abu-Mostafa and Atiya, 1996) will have a direct bearing on future decision making such as fresh investment on, sale or status-quo of the stocks.

Despite proliferating research in the field, forecasting future stock prices is a complicated process since stock market exhibits the dynamic trend. It is all the harder if we want to forecast stock price based on relevant news articles. It is well-known that the raw text data is not useful for any data mining task. Therefore, we convert the text into an intermediate form called Document-Term Matrix using which one can perform syntax-based document classification based on some tokens or features. But, in sentiment/ opinion mining tasks these syntactical features do not play a significant role in knowledge acquisition. Semantic features are helpful for understanding the customer behavior/ opinion analysis. Such that Semantic features play a vital role in sentiment analysis. Extraction of the semantical feature, where each feature maps an opinion word poses a significant challenge. There are various methods available for extraction of semantic/ linguistic features. Few of them are, OpinionFinder, Linguistic Inquiry and Word Count (LIWC), Google Profile of Mood States (GPOMS), SentiWordNet, R sentiment analysis and Python NLP package, etc.  The features extracted by these tools are based on opinions, writing style and mood in which a particular article was written. This approach alleviates the problem faced by syntactic features in not being able to present the hidden semantic meaning of the text that can represent a real pattern. However, LIWC & TAALES software extract psycholinguistic features unlike other methods mentioned above.

In literature, there is a huge number of stock prediction models that deal with only numeric data under time series analysis framework and comparatively not much work is reported in stock prediction using text mining of financial news articles. So far, models were built using only news headlines that have limited text with no details of the entire information. It is evident from the news that the news article contains more details instead of news headlines. Therefore, the sentiment of a news article can be a useful predictor for forecasting a stock. Therefore, in this paper, an attempt is made to predict the stock price of a company using the information contained in news articles related to the particular company in question.  We conjecture that a correlation exists between news and the stock values.  The sentiment present in news articles contains useful information about stock price forecasting.

The contributions of this paper are:

1. Extraction of psycholinguistic features from the financial news articles concerning Indian companies. These features collectively convey the sentiment hidden in the article.
2. Extraction of lexical sophistication features from financial news articles.
3. Imputing missing linguistic/ lexical feature values for the cases where the stock price is available for a company but the corresponding news is not.
4. Developing stock prediction models with these features as predictor variables using a host of intelligent techniques.

The structure of the rest of the paper is as follows. Section 2 reviews the works related to text mining in stock market analysis. Section 3 describes the methods applied in current work. Section 4 presents the proposed methodology. Section 5 presents the results of our analysis and comparative analysis of models. Finally, we summarize our work by concluding it in Section 6.

## 2. Motivation for the present work

(i) A Majority of the existing works in literature initially categorized the news articles and later performed the prediction tasks. But, in our approach, we predicted stock value based on news articles.

(ii) Unlike the extant studies that employ conventional sentiment analysis tools, we wanted to extract psycho-linguistic features from the news articles and use them as predictor variables to predict stock price.

(iii) We employed Linguistic Inquiry and Word Count (LIWC) and TAALES tools for extracting psycho-linguisti, lexical feactures. LIWC provides 93 types of psycho-linguistic features, TAALES provides 241 features whereas other tools such as GPOMS provide few expressions/ linguistic features like six mood states namely Alert, Calm, Happy, Kind, Sure and Vital, while Sentiwordnet consists of positive or negative opinions only. To the best of our knowledge, this proposed approach is hitherton ever reported in the literature for predicting the Indian firms' stock prices.

## 3. Literature Review

Financial markets drive a lot of investment decisions all over the world. Stock markets witness dramatic changes over time in response to the geo-political, social and fiscal changes globally. These in turn trigger financial risks in investments with the investors and the financial institutions being the stakeholders. Consequently, researchers started studying the cause and effect relationship between various market factors and the corresponding movements in stock prices. Most of the works focused on quantitative data like historical/ actual prices as predictor variables to predict the present stock price. Less attention was paid to the use of the enormous amount of unstructured textual data generated from the web in the form of published news articles, public opinions in social media and blogs by experts in the field of financial investments.

In this section, the past works of investment risk modeling and market predictions using this unstructured data is briefly reviewed.

Engle et al., (1993) predicted volatility in the stocks using news. Autoregressive Conditional Heteroskedasticity (ARCH), Generalized Autoregressive Conditional Heteroskedasticity (GARCH) models are fitted on stock returns of Japan from 1980 - 1988. They concluded that impact of volatility in the negative news is higher than positive news for stock returns. Lavrenko et al., (2000) presented a model to identify the news stories which affect the trend of financial markets. They identified the patterns in the time series with the help of piecewise linear fit followed by label assignment with an automated binning process. They concluded that particular stock related news is useful for analysis compared to the global news.

Thomas and Sycara (2000) worked on the behavior of financial markets. Textual information is available on the website of a company impacts its business. They proposed two models based on maximum entropy and genetic algorithm to predict financial markets. They concluded that the combination of these two models outperformed the stand-alone models.

Then, Peramunetilleke and Wong (2002) proposed a new model for forecasting exchange rates based on the current status of world financial markets. The study investigated on how news headlines of the financial market could be helpful for forecasting the currency exchange rates.

They concluded that the proposed approach was better than random guessing and suggested that hybrid models for better prediction.

Koppel and Shtrimberg (2006) proposed a model based on the news articles for stock prediction. They extracted the features from the Multex Significant Development corpus and predicted the Standard & Poor 500 (S&P 500) stock index. During the process of modeling, they labeled the news as positive or negative according to their impact on the price. Later, they employed SVM to train the news articles and reported an accuracy of 70%.

Rachlin et al., (2007) proposed a model called ADMIRAL, based on textual information of web documents and time series data. They employed automatic extraction of text instead of the predefined expert list. They explained the functionality of ADMIRAL which consists of six steps: Data collection, feature extraction, term weighting, and combined data construction, classification using decision tree (DT) and market recommendation. They acquired the data from the online sources of Forbes and Reuters. They reported an accuracy of 83.3% with DT on both the datasets.

Zhai et al., (2007) presented a model for stock price prediction using news and technical indicators as explanatory variables. They employed SVM for classification. They considered the daily share prices of BHP Bilton Ltd. from Australian Stock Exchange as output. Their method yielded higher directional prediction accuracy of 70% compare to specific models considering news alone or technical indicators alone.

Mahajan et al., (2008) analyzed the impact of news on the stock market. They identified the events by employing Latent Dirichlet Allocation (LDA) based topic extraction method. They analyzed the actual market data with news to understand the impact on the SENSEX market. They developed a hybrid model by combining the DT and SVM and reported a prediction accuracy rise or fall of 60%.

Evans and Lyons (2008) also experimented with macro news for studying the currency flow. In this work, they observed that the arrival of macro news could account for more than 30% of daily price variance. They experimented with US News and FX rates with standard error as a performance metric. They concluded that macro news impacted two-thirds of the directional movement/ exchange rates.

Butler and Keselj (2009) presented a model based on N-gram analysis for financial forecasting. They constructed various models using character n-gram, word n-gram, a hybrid of readability model with SVM and hybrid of readability and n-gram. They predicted closing values of the S&P 500 companies with the help of the textual information present in their annual reports. They concluded that hybrid model of character n-gram yielded the best performance compare to other models concerning the percentage of returns.

Bollenet al., (2010) proposed a model for stock prediction using Twitter tweets. The opinion of the tweets are extracted using OpinionFinder and GPOMS tools, where Opinion Finder consists of positive or negative opinion, GPOMS consists of 6 mood states namely alert, calm, happy, kind, sure, and vital. They employed a self-organizing fuzzy neural network for prediction of the Dow Jones Industrial Average values. They predicted the up and down values of the stock (closing values) with an accuracy of 87.6%. They concluded that through this approach the MAPE value was reduced by more than 6%.

Groth and Muntermann (2011) published work in the field of intra-day market risk management by using textual data analysis to discover patterns that can explain risk exposure. Different learners used included Naive Bayes, k-Nearest Neighbor, Neural Network, and Support Vector Machine to processed feature datasets followed by traditional measures of evaluation namely accuracy, recall, precision and F-measure as well as domain specific simulation-based model evaluation. The results clearly supported the influence of textual information in financial risk management.

Chan and Franklin (2011) proposed a novel text-based decision support system which extracts event sequences from text patterns and predicts the likelihood of the occurrence of events using a Hidden Markov Model -based inference engine. They investigated more than 2000 financial reports with 28,000 sentences. Experiments showed that the prediction accuracy of the model outperformed similar statistical models by 7% for the seen data while significantly improving the prediction accuracy of the unseen data. Further comparisons substantiate the experimental findings.

Li et al., (2011) proposed a model for stock market prediction by integrating quantitative and qualitative information. They collected the news articles during the Hong Kong Stock Exchange trading time. After pre-processing of text, they generated tf-idf matrix and applied Chi-square feature selection method to find out prominent features. They employed NaiveBayes (NB), Multi-Kernel Learning (MKL) and SVM techniques. They concluded that MKL outperformed other models.

Vu et al., (2012) proposed a model for predicting stock price up and down movements based on Twitter messages. Initially, they labeled the sentiment into two categories – positive and negative. Based on this, they predicted the stock price of four companies' viz., Amazon, Apple, Microsoft, and Google with 41 days' data using Decision Tree. Reported the accuracies values are 75%, 82.93%, 75.61% and 80.49% respectively.

Hagenau et al., (2013) proposed the use of robust feature selection approaches for stock prediction. Chi-square and bi-normal separation to select semantically relevant features, to improve classification accuracy for financial stock prediction. Initially, they classified the news articles later, they constructed the prediction model. They experimented on German, UK announcement with stock values available in Data stream. They built the various models with various feature subset selection methods viz., single words, 2-Gram, 2-word combination using SVM. They concluded that with 2-word combinations they reported an accuracy of 76%.

Jin et al., (2013) proposed a model called for Forex-foreteller which mines news articles and forecasts the movement of foreign currency markets. A combination of language models, topic clustering, and sentiment analysis was used to identify the relevant news articles. These were combined with historical stock index and currency exchange values for prediction. They employed linear regression model for currency forecasting. They experimented with the Argentina, Brazil, Chile and Columbia currencies concerning US Dollar value. They concluded that with this proposed model they reported higher recall values of 0.6, 0.63,1 and 1 respectively compare to precision values.

The effect of macro news on upward and downward movements of FOREX is studied by Chatrath et al., (2014). They employed multivariate regression model in this approach. They investigated the currencies of UK, Japan, Swiss and Euro onthe arrival of news. They observed that US announcements are directly linking towards nearly of 15% currency jumps. They concluded that 56% of currency change is happening within the 5 min of news arrival.

Li et al., (2014) presented work based on Extreme Learning Machine (ELM) for stock market prediction. By considering the news articles and stock prices, they employed SVM and Neural Network, etc. They carried out the experiments on 23 stocks of the H-share (Chinese) market and its corresponding news. They concluded that with the proposed ELM approach outperformed other techniques.

FOREX market prediction using news headlines as predictors is reported by Nassirtoussi et al., (2015). They proposed a multilayer architecture consisting of semantic abstraction, sentiments aggregation, and dynamic model creation. They concluded that their approach yielded an accuracy of 83.33%.

Shynkevich et al., (2016) presented a presented a framework to find out the stock movements using Multiple Kernel Learning (MKL). They extracted the news from LexisNexis source and categorized the news based on their relevance to the stock, industry, and sub-industries, etc. After preprocessing, they applied chi-square method for feature selection on tf-idf matrix. For experimental purpose, they considered S&P 500 index stocks in Health sector. They employed SVM, k-NN, and MKL techniques. They concluded that (i) the predictive performance of all models are better due to the various types of news sources. (ii) Proposed MKL method performed better than other two models.

According to the behavioral economics moods, sentiment and emotions are playing a significant role in investors' decision-making process. Ho and Wang (2016) developed a model for predicting stock market movement using Artificial Neural Network (ANN). They experimented on stock prices of Google (NASDAQ:Google) and News articles of Dow Jones for predicting upward and downward movement of the stock. They evaluated the model with prediction rate, sensitivity, and specificity. They concluded that the proposed model is better than Random walk forecast method.

It is evident from the overview of past works in stock market prediction is that various information sources are combined to produce a joint feature set. It may not provide valid information for assessing the effect of each source on the stock. To overcome this difficulty, Li et al., (2016) proposed a framework using Tensor methods for stock market prediction. Through this approach, they could capture the essential information among multiple sources. They experimented with the data sources like CSI 100 stocks, financial discussion boards, and news reports. The performance evaluation was carried out with Directional Accuracy (DA) and Root Mean Square Error (RMSE) values.

## 4. Overview of methods applied

In this section, we describe the two feature selection methods employed followed by various data mining algorithms.

### 4.1. Feature selection methods

Feature subset selection is an important task in any text mining task. In this work, we used the following two feature subset selection methods.

#### 4.1.1. Chi-square method

Chi-squared helps us decide whether a categorical predictor variable and the target class variable are independent or not. High chi-squared values indicate the dependence of the target variable on the predictor variable. It is employed in many text mining applications (Zheng et al., 2004).

#### 4.1.2. Minimal Redundancy and Maximal Relevance (MRMR)

Minimum redundancy maximum relevance (MRMR) (Peng et al., 2005) feature selection method uses a heuristic to minimize redundancy while maximizing relevance to select promising features for both continuous and discrete datasets. The maximum relevance condition is obtaining by through features F-statistic values. For further details, the reader is referred to Peng et al., (2005), Ding and Peng (2005).

## 4.2. Machine Learning Techniques

In this work, we employed the following machine learning algorithms.

### 4.2.1. Support Vector Regression (SVR)

Support Vector Machines (SVM) (Cortes and Vapnik, 1995) proved useful for solving classification problems. However, Support Vector Regression (SVR) (Gunn, 1998) uses the same methodology as that of SVM barring few changes to solve regression problems. SVR is employing in various applications like power consumption estimation, financial market forecasting (Yang, 2002), electricity price (Sansom et al., 2002), travel time prediction (Wu, Ho, & Lee, 2004) and software cost estimation (Pahariya et al., 2009).

### 4.2.2. Random Forest (RF)

Ho (1995) proposed Random Forest. It builds multiple trees on a randomly selected feature subset on a sample of data obtained with replacement (also known as bootstrap sampling). It is expandable for increasing the performance on both training and test data. It performs both classification and regression and also handles higher dimensions of the datasets.

### 4.2.3. Quantile Regression Random Forest (QRRF)

Quantile Regression Random Forest was introduced by Meinshausen(2006). The significant difference between RF and QRRF is as follows: All observations are kept in a node in quantile regression random forest whereas in the random forest node contains the mean of observations only. It is just like an optimization problem i.e. conditional mean estimation is performed by minimizing the squared error so that quantiles reduce the expected loss. Selection of suitable parameters for quantile regression minimizes the empirical loss. The quantile regression random forest is non-parametric and yields accurate predictions. In this connection, Ravi and Sharma (2014) proposed a hybrid model using SVR and QRRF in tandem for regression tasks.

### 4.2.4. Classification and Regression Tree (CART)

CART is proposed by Breimanet al., (1984). It is one of the decision tree algorithms that solves both classification and regression problems. It has the following advantages: automatic variable selection, handling missing values, handling discrete as well as continuous variables. In this algorithm, the splitting of the root node is based on the sum of squared errors. It is too popular to be described here.

### 4.2.5. Multilayer Perceptron (MLP)

It is the most popular neural network model that maps a set of input variables onto a set of output or target variables. It contains an input, hidden, and an output layer. Hidden layer explains the nonlinearity of the dataset. MLP uses a standard back propagation algorithm to estimate the weights connecting these layers. MLP is a universal approximator and is widely used for solving both classification and regression problems (Rumelhartet al., 1986).

### 4.2.6. Data Handling (GMDH)

Group Method Data Handling (Ivakhnenko, 1968)is the first deep learning neural network in a broad sense with several hidden layers and is using in various applications such as pattern recognition, forecasting, and systems modeling. It is using in different applications like energy demand prediction (Srinivasan, 2008), bankruptcy prediction (Ravisankar and Ravi, 2010), software reliability prediction (Mohantyet al., 2013), credit card churn prediction (Sundarkumar and Ravi 2015), software cost estimation (Pahariyaet al., 2009), insurance fraud detection (Sundarkumar and Ravi 2015), phishing detection (Pandey and Ravi, 2013), web service classification (Mohantyet al., 2010), forecasting FOREX rates (Pradeepkumar and Ravi, 2014), etc. There is an advantage with GMDH is that it automatically selects the number of hidden layers and neurons in each hidden layer. The network is thus composed of active neurons that organize themselves. The GMDH network learns in an inductive way and tries to build a polynomial function which minimizes the error between the predicted value and expected output. For more details, the reader can refer to Ivakhnenko (1968).

### 4.2.7. General Regression Neural Network (GRNN)

Specht (1991) proposed GRNN. It is useful to solve regression problems and it contains four layers namely input, pattern, summation and output layers in that order. It has the following features: quick learning, easy training, and outlier discrimination. It can approximate any function from the past data. It simply implements the non-parametric regression to find the best fit for the observed data. GRNN is widely used in various applications including FOREX rate prediction (Pradeepkumar and Ravi, 2014), Software Reliability Prediction (Mohanty et al., 2013).

## 5. Proposed Methodology

In this work, a hybrid model which performs text mining on the financial news articles and forecasting of the stock price in tandem is proposed. The proposed methodology consists of three phases namely preprocessing, imputation and forecasting as depicted in Figure 1. Initially, all news articles and the corresponding stock prices of a set of companies were collected. Datasets' description can be found in the Section 6. Later, the news articles were preprocessed by employing LIWC (2015) and TAALES (Kyle et al., 2017) software. The output, i.e., the documents and their corresponding set of linguistic feature values, was captured in a structured format. Stock value, the target variable, was appended to the matrix obtained by using LIWC tool, where the columns of the matrix were used as predictor variables. Details about the linguistic features and LIWC are presented in Section 5.1. Same process was repeated with the features extracted using TAALES. LIWC and TAALES was recently employed by Ravi and Ravi (2017) for irony and satire detection in news and textual corpora. In imputation phase, initially the missing records i.e., examples where the stock value is present, but corresponding news articles respect to the particular stock is not available were found. Then, the neighbors of a stock value within a range of 10% were selected, and imputation was performed with respect to stock value using mean. In the final phase, i.e., modeling, initially the regression models were built with all features. Later, two feature selection methods namely Chi-square and MRMR were applied for identifying the discriminative features. Finally, experiments were conducted with

top-10 as well as top-25 feature subsets. Finally, MAPE and NRMSE values were reported for performance evaluation.

## 5.1. Linguistic Features

We employed the Linguistic Inquiry and Word Count (LIWC) to find out the linguistic features in the news articles. LIWC includes (Tausczik and Pennebaker, 2010) the text analysis module along with a group of built-in dictionaries which is used to count the percentage of words reflecting different emotions, thinking styles, social concerns, and even parts of speech. LIWC counts the words which are in psychologically meaningful. It contains a dictionary of 6400 words (Pennebaker et al., 2015). These words are again containing sub-dictionaries. The output of LIWC contains 93 variables; these variables may be belonging to the following groups: General description, the summary of language, linguistic, personal concern, physiological, personal.

## 5.2. Lexical Sophistication Features

TAALES (Tool for the Automatic Analysis of Lexical Sophistication) was employed to extract lexical features from the news articles. TAALES (Kyle and Crossley, 2015) calculates various indices, and these are related to the frequency of the words, and its ranges, n-gram frequency, word neighbors, strength association between the words, psycholinguistic properties of the words, word recognition norms (standard deviation), polysemy, Mutual information, etc. There are two versions of this software viz., TAALES 1.0 and TAALES 2.0 (Kyle et al., 2017). TAALES 1.0 consists of 103 indices, whereas TAALES 2.0 consists of 424 indices. The output of TAALES 2.0 contains 241 variables. These indices are created from the British National Corpus.

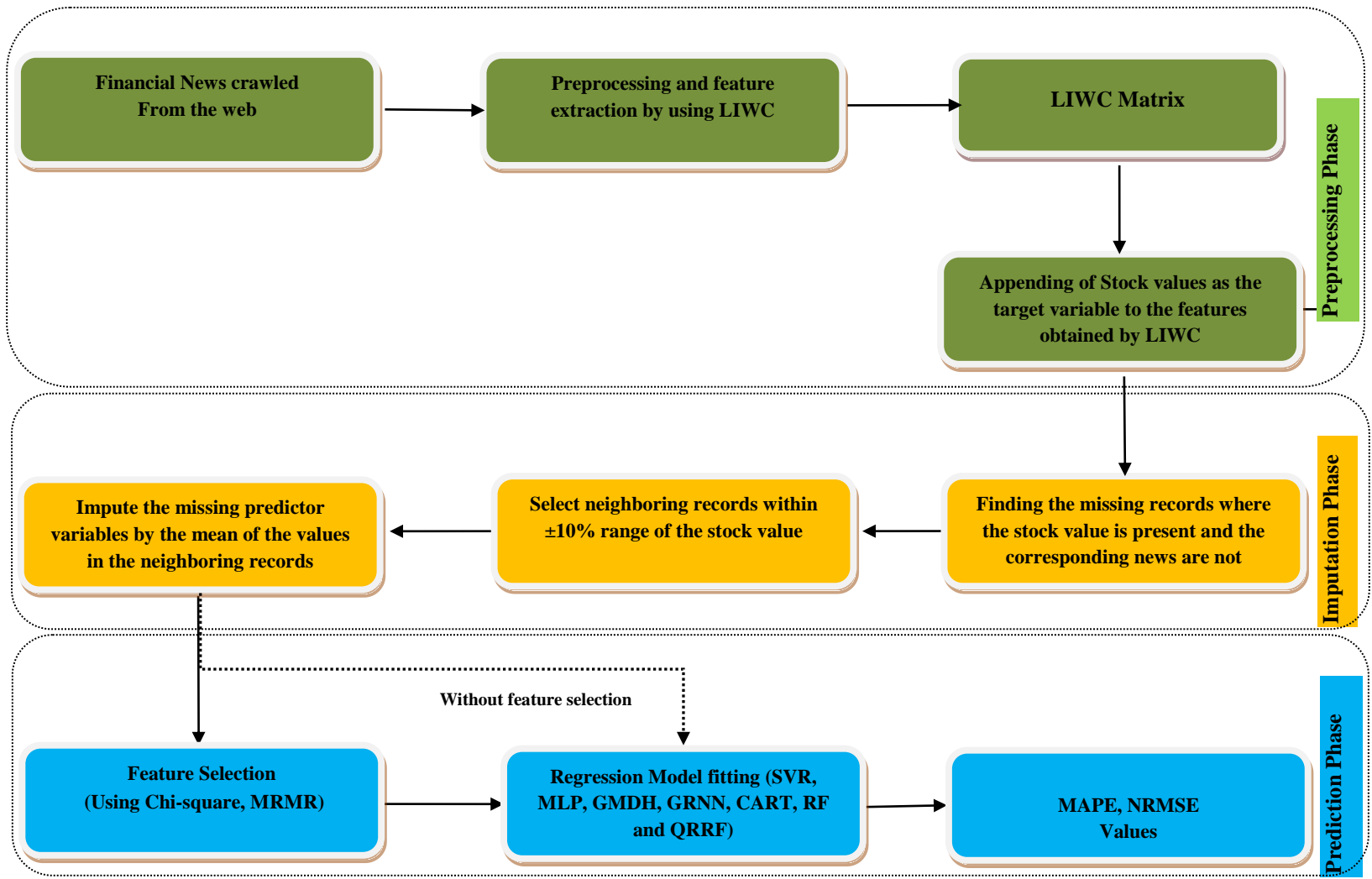

**Fig. 1 Proposed Methodology**

# 6. Experimental Design

## 6.1. Data acquisition

The data used in the proposed experiments was collected through "Business Standard" online news resource. This involved collecting news articles for 12 major Indian companies including, Bharti Airtel Limited, Mahindra & Mahindra Limited, Tata Consultancy Services Limited (TCS), Tata Motors Limited, Reliance Industries Limited, Tata Steel Limited, State Bank of India (SBI) and Oil and Natural Gas Corporation (ONGC). The corresponding historical stock prices were extracted from the "Yahoo Finance India" online web resource (Yahoo Finance, 2016). The tools used for web crawling were "Web Scraper" (Web Scraper, 2016). The description of datasets is outlined in Table 1.

## 6.2. Dataset description

To validate our proposed method, we conducted novel experiments with the following data sets, extracted from the web. The description of the datasets is listed in Table 1.

**Table 1.** Distribution of news articles with respect to the company

| S.No. | Dataset | Number of articles | Period of data availability |
|---|---|---|---|
| 1 | Bharti Airtel | 85 | 28th January 2016 to 25th May 2016 |
| 2 | Mahindra | 125 | 1st December 2015 to 31st May 2016 |
| '23 | Tata Motors | 108 | 26th November 2015 to 22nd April 2016 |
| 4 | Reliance Industries | 116 | 29th November 2015 to 9th May 2016 |
| 5 | Tata Steel | 131 | 2nd December 2015 to 30th May 2016 |
| 6 | TCS | 126 | 10th December 2015 to 27th May 2016 |
| 7 | SBI | 126 | 7th December 2015 to 30th May 2016 |
| 8 | ONGC | 125 | 22nd December 2015 to 31st May 2016 |
| 9 | Infosys | 130 | 22nd December 2015 to 15th June 2016 |
| 10 | Sun Pharma | 129 | 11th December 2015 to 5th June 2016 |
| 11 | Spice Jet | 131 | 21st December 2015 to 24th May 2016 |
| 12 | Jet Airways | 125 | 14th December 2015 to 3rd June 2016 |

It has been a general observation that news articles are not published every day for every company in the Indian Stock markets. But on the other hand, Stock prices are available for all days except the weekends (Saturday and Sunday) and national holidays. So the missing values of all predictor variables in the entire record were imputed using the method described below.

### 6.3. Data imputation process

In today's world, handling incomplete data is a very common difficulty in most of the datasets. There are various causes for missing data: weak data acquiring process, data privacy issues, non-availability of data and many other reasons. It leads to uncertainty in the dataset and causes inaccurate prediction. So, here imputation plays a significant role.

Imputation is defined as the process of replacing missing values with substituted values. Imputation plays a significant role in datasets in various fields, including financial, speech processing and medical diagnostics, etc. In a dataset, it is necessary to know the reason for missing data. There are various types of missing data. Little and Rubin (1987) categorized the missing data into three categories namely

1. MCAR (Missing Completely at Random): According to MCAR, the missing data mechanism is unrelated to values of any other variable in the dataset.

2. MAR (Missing at random): MAR mechanism is involved when the probability of missing values corresponding to a particular variable is related to some other variable in the dataset but not with the variable itself.

3. MNAR (Missing Not at Random): According to MNAR, the missing values on a variable are related to the variable itself and not on other controlled variables in the dataset.

Conventional imputation methods include mean imputation and regression imputation. Multiple imputation methods involve replacing missing value with a set of plausible values. These imputed datasets are analyzed by using standard procedures. Nishanth and Ravi (2016) proposed mean imputation followed by running probabilistic neural network for imputation and tested its effectiveness on a set of benchmark problems. Earlier, Gautam and Ravi (2015) proposed two models for data imputation based on Counter Propagation Auto-Associative Neural Network (CPAANN) and Grey System Theory with CPAANN. Then, Ravi and Krishna (2014) proposed a hybrid model for data imputation using mean imputation followed by General Regression Auto Associative Neural Network (GRANN) or Particle Swarm Optimization based Auto Associative Neural Network (PSOANN). They concluded that GRANN outperformed other models on four benchmark datasets. However, we employed an imputation method different from the above.

News articles corresponding to a particular company are not published every day. It creates gaps in the time series values of the LIWC/ TAALES feature scores. Hence, in this scenario data imputation plays a significant role. Before we present the imputation procedure, some data preprocessing steps employed in this work are noteworthy. In some cases, it was found that multiple news articles were available for a particular company on the same day. Consolidated LIWC/ TAALES feature scores for that date could be calculated by averaging the individual LIWC/ TAALES scores of all the news articles published on that day. Further, it is a known fact that stock markets remain closed on weekends i.e. Saturday and Sunday, and on public holidays. This leads to a situation where news articles are available on a particular day when there is a stock market holiday. Losing this data will lead to information loss. To retain such feature values, all these data instances with no available stock prices on the corresponding date are merged into next instance with available stock price. This is done by averaging out values of all these instances till the date where next stock price information is available.

In this approach, initially the records where the stock value is available, and the corresponding news is not available or missing is found. For every record with a given stock price and missing financial news, the missing feature values are imputed as follows:

(i) Pick up all those records whose stock price is within plus or minus 10% of the current stock price.

(ii) Compute the mean of all the feature values of the records so chosen in order to form a new feature vector.

(iii) Finally, this feature vector acts as a proxy for the missing financial news.

(iv) For imputation, the method of Ling and Mei (2009) was adopted.

Similar works are also found in the literature on imputation (Patilet al., 2010), Garcia-Laenciana et al., (2009k). The missing values are finally imputed using the following formula..

$$x_i' = \sum_{i=1}^{n} \frac{x_i}{d_i} , \quad \forall x \text{ (every attribute of the corresponding stock value)}$$

*where $x_i$' is the imputed value for the missing $i^{th}$ attribute, $x_i$ is the attribute value obtained in the step (ii) above and $d_i$ represents the absolute difference between the corresponding stock value of missing record and that obtained in step (ii) above.*

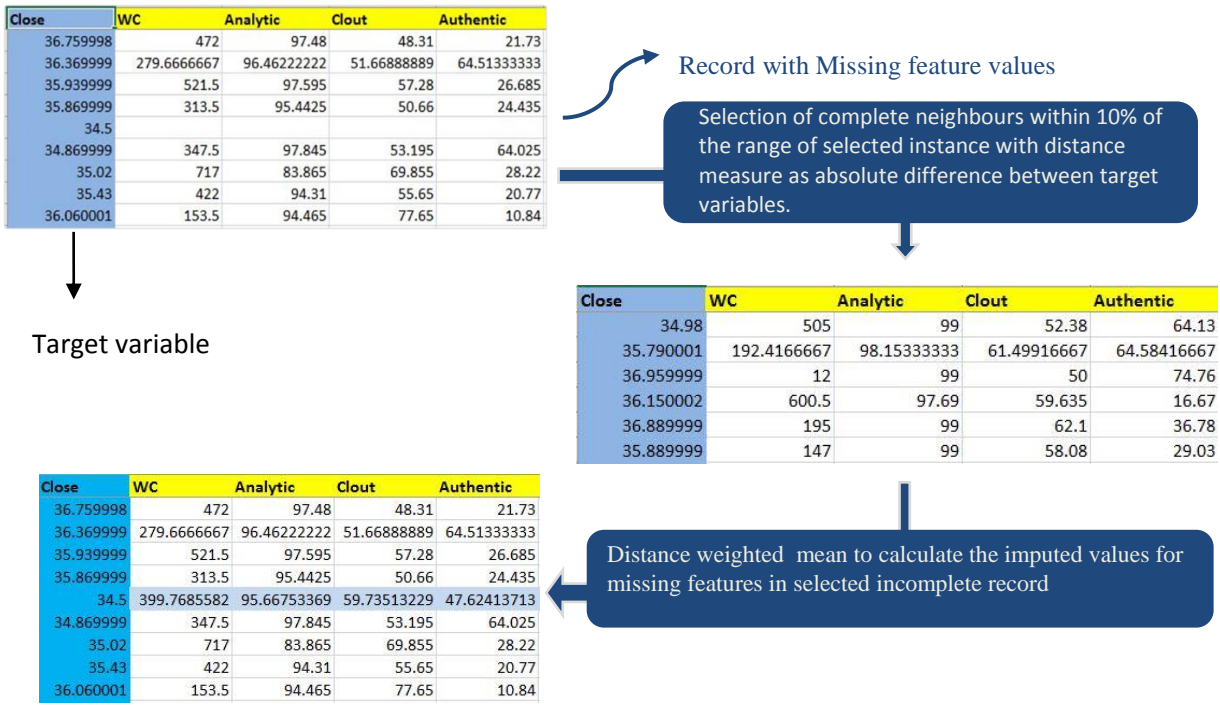

**Fig. 2** Data Imputation process

## 6.4. Experimental Procedure

All the experiments are conducted on a computer having $i_5$ processor with 2.6GHz, 8GB RAM, 500GB HDD and 64-bit operating system of Windows 8. We used R language packages (2014) for RF, QRRF, RPART and SVR. We employed GMDH, GRNN models using Neuroshell (2010). Similarly, for MLP we used Statistica Trial Version (2016). We presented the various parameter settings for distinct models in Table 2.

## 6.5. Performance Measures

In this paper, we evaluated the performance of the proposed system with the following metrics. Mean Absolute Percentage Error (MAPE) (Flores, 1986) and Normalized Mean Square Error (NRMSE). The stock value varies from one company to another company. Hence haven't considered/ reported Mean Squared Error (MSE) value as a performance metric. For uniform scaling, we reported these values only.

$$MAPE = \frac{1}{n}\sum_{i=1}^{n}\frac{|y_i - \hat{y}_i|}{y_i} * 100$$

$$NRMSE = \frac{1}{Range}\sqrt{\frac{\sum_{i=1}^{n}(y_i - \hat{y}_i)^2}{n}}$$

*$y_i$=Actual Value, $\hat{y_i}$=Predicted Value, n=number of observations, Range = max − min*

**Table 2. Parameter settings for various models**

| RPART | |
|---|---|
| Parameter | Value |
| Minsplit | 20 |
| minbucket i.e. round(minsplit/3) | 7 |
| complexity parameter (cp) | 0.01 |
| Maxcompete | 4 |
| maxsurrogate | 5 |
| Usesurrogate | 2 |
| Maxdepth | 30 |
| **SVM** | |
| Type | eps-regression |
| Kernel | radial |
| Cost | 1 |
| Epsilon | 0.1 |
| no. of support vectors | 78 |
| **Random Forest** | |
| ntree | 1000 |
| node size | 5 |
| maximum nodes | 83 |
| **QRRF** | |
| No. of trees | 200 |
| No. of variables used for split | 5 |
| **MLP** | |
| Hidden units | 4 |
| Max hidden units | 13 |
| Networks to train | 20 |
| Networks to retain | 5 |
| Error function | Sum of squares |
| Activation function | Tanh |
| Cycles | 200 |
| Learning rate | 0.1 |
| Momentum | 0.1 |
| **GMDH** | |
| Scale function | Linear(-1,1) |
| type | advanced |
| Maximum variables in connection | $x_1 x_2 x_3$ |
| Maximum product terms in connection | $x_1 x_2 x_3$ |
| Max variable degree connection | $x^3$ |
| Selection criterion | GCV |
| Type of schedule | asymptotic |
| Optimization of the model | full |
| missing values treated as | error |

| GRNN | |
|---|---|
| Smoothing factor | 0.3 |
| Scaling function | Linear[0,1] |
| distance | Vanila(euclidean) |
| caliberation | Genetic, adaptive |
| Genetic breeding pool size | 300 |
| Auto termination of the generations with no improvement of 1% | 20 |
| Missing values treated as | error |

## 7. Results and Discussions

The Linguistic Inquiry and Word Count (LIWC) software was employed to find out the linguistic features in the news articles. LIWC includes (Tausczik and Pennebaker 2010) a text analysis module along with a group of built-in dictionaries which is used to count the percentage of words reflecting different emotions, thinking styles, social concerns, and even parts of speech. LIWC counts the words which are psychologically meaningful. It contains a dictionary of 6400 words (Pennebaker et al., 2015). These words again contain sub-dictionaries. The output of LIWC contains 93 variables; these variables belong to one of the following groups: General description, the summary of language, linguistic, personal concern, physiological, personal. Similarly we employed TAALES for extracting lexical sophistication features. The output contains 241 variables. The results of all models with LIWC and TAALES features are presented in the following cases,

We presented the results of all models in various Cases viz., (i) Full features, (ii) Chi-square top-25 features (Ch-25), (iii) Chi-square top-10 features (Ch-10), (iv) MRMR top-25 features (MRMR-25) and (v) MRMR top-10 features (MRMR-10) .

For all datasets, the results with LIWC features are presented in Table 3 through Table 7. In these tables, cells highlighted in */ green indicate the best performance of the model under consideration vis-à-vis other models.

**Table 3. Stock Prediction Results with All Features from LIWC Features**

| Dataset | GMDH | | GRNN | | RF | | QRRF | | RPART | | SVR | | MLP | |
|---|---|---|---|---|---|---|---|---|---|---|---|---|---|---|
| | MAPE | NRMSE | MAPE | NRMSE | MAPE | NRMSE | MAPE | NRMSE | MAPE | NRMSE | MAPE | NRMSE | MAPE | NRMSE |
| Airtel | **0.067** | **0.014** | 1.150 | 0.48 | 4.89 | 0.806 | 3.11 | 0.510 | 5.56 | 0.925 | 9.69 | 0.89 | 2.77 | 0.513 |
| Mahindra | **0.61** | **0.025** | 9.49 | 0.36 | 16.99 | 0.617 | 13.31 | 0.469 | 17.99 | 0.641 | 15.66 | 0.565 | 7.709 | 0.375 |
| Tata Motors | **0.367** | **0.045** | 0.485 | 0.051 | 7.111 | 0.593 | 5.153 | 0.455 | 9.108 | 0.809 | 6.66 | 0.576 | 6.155 | 0.557 |
| Reliance Industries | **0.182** | **0.026** | 0.269 | 0.038 | 2.589 | 0.355 | 2.070 | 0.284 | 3.006 | 0.415 | 2.196 | 0.314 | 2.723 | 0.347 |
| Tata Steel | **0.422** | **0.039** | 0.496 | 0.051 | 12.82 | 1.030 | 7.47 | 0.662 | 10.159 | 1.086 | 10.247 | 0.921 | 8.986 | 0.817 |
| TCS | **0.136** | **0.0378** | 0.524 | 0.169 | 3.37 | 0.878 | 3.26 | 0.780 | 4.72 | 1.371 | 3.22 | 0.841 | 2.610 | 0.727 |
| SBI | **0.292** | **0.024** | 0.533 | 0.045 | 4.954 | 0.371 | 3.206 | 0.239 | 9.106 | 0.691 | 4.855 | 0.374 | 5.45 | 0.360 |
| ONGC | 0.360 | 0.067 | **0.209** | **0.068** | 1.446 | 0.235 | 0.829 | 0.152 | 2.329 | 0.443 | 1.21 | 0.19 | 1.780 | 0.298 |
| Infosys | **0.131** | **0.024** | 0.846 | 0.186 | 3.52 | 0.597 | 3.11 | 0.536 | 2.785 | 0.475 | 3.553 | 0.621 | 2.113 | 0.411 |
| Sun Pharma | **0.104** | **0.012** | 0.527 | 0.088 | 2.472 | 0.317 | 1.986 | 0.284 | 2.779 | 0.333 | 2.715 | 0.358 | 3.057 | 0.360 |
| Spice Jet | **0.192** | **0.007** | 0.215 | 0.009 | 3.385 | 0.173 | 2.134 | 0.152 | 4.338 | 0.213 | 2.709 | 0.146 | 0.705 | 0.045 |
| Jet Airways | 0.318 | 0.028 | **0.244** | **0.019** | 3.195 | 0.279 | 2.428 | 0.214 | 6.485 | 0.533 | 2.598 | 0.216 | 4.549 | 0.332 |

Table 3 presents the results in terms of MAPE and NRMSE corresponding to various prediction models without feature selection using LIWC features. GMDH outperformed all other techniques in terms of both MAPE and NRMSE, in all but two datasets (ONGC, Spice Jet). For these datasets, GRNN performed well.

Table 4 presents the results of various models fed by the top-25 features obtained by Chi-square feature selection method using LIWC features. It can be observed from the table that GMDH yielded the best predictions in all datasets except SBI, ONGC, SpiceJet and Jet Airways. For these four datasets, GRNN outperformed all other techniques. It is to be noted that in case of Mahindra, Tata Motors and Reliance Industries datasets, Chi-square value returned is 0 for most of the features. It means that they have no impact on prediction. Therefore, the results of these datasets are not presented in Table 4.

Table 5 presents the results of the models with top-25 features selected by MRMR method using LIWC features. In this combination, GMDH performed the best in terms of MAPE and NRMSE on all datasets except SBI, ONGC, Infosys, Sun Pharma, Spice Jet and Jet Airways datasets; for these datasets, GRNN performed the best. Table 6 presents the results of the models trained with top-10 features selected by Chi-square method using LIWC features. From this table, it can be observed that GMDH outperformed all other techniques on the datasets of Airtel, Mahindra, Tata Motors (8 features), TCS, Infosys, Sun Pharma. Whereas, GRNN could yield the best predictions on Reliance Industries, Tata Steel, SBI, ONGC, Spice Jet, and Jet Airways in terms of both MAPE and NRMSE values. Interestingly, GMDH and GRNN performed almost identically on Tata Steel.

Table 7 presents the results with top-10 features selected by MRMR method using LIWC features. The table shows that GRNN outperformed other models in terms of both MAPE and NRMSE on seven companies' stocks (Tata Motors, Reliance Industries, SBI, ONGC, Infosys, SpiceJet, and Jet Airways) and GMDH performed the best on the remaining five datasets.

Further, the features (LIWC) selected through Chi-square and MRMR methods are presented in Table 8. The models that are not statistically significant compared to case (i) (full features case) are only reported here. From Table 8, it can be inferred that the psycholinguistic features having highest frequency of occurrence across different data sets are as follows: achieve, Analytic, male, relig, Comma and QMark.

The excellent performance of GMDH in most of the datasets is attributed to the fact that it is one of the earliest Deep learning neural networks thereby possessing very high predictive power. The non-parametric regression, which is at the heart of the GRNN does the trick for its second best performance behind GMDH. In order to determine the usefulness of feature subset selection methods employed here with LIWC features, we conducted a statistical significance test called Diebold-Mariano Test (DM) Test (Diebold and Mariano, 2002) between Case (i) of GMMDH (LIWC) and all other cases of GMDH (LIWC) in a pair-wise manner for all datasets except ONGC and Jet Airways. For these 2 datasets, DM test was performed between Case (i) of GRNN (LIWC) and all other cases of GRNN (LIWC) in a pair-wise manner. GMDH and GRNN were chosen because of their superior performance over other models in terms of MAPE and NRMSE as seen in Tables 3 through 8. The code for the test is available at DM Test (2017). The DM Test values are reported (with LIWC features) in Table 9.

**Table 4. Stock Prediction Results with Chi-square (25) feature selection method**

| Dataset | GMDH | | GRNN | | RF | | QRRF | | RPART | | SVR | | MLP | |
|---|---|---|---|---|---|---|---|---|---|---|---|---|---|---|
| | MAPE | NRMSE | MAPE | NRMSE | MAPE | NRMSE | MAPE | NRMSE | MAPE | NRMSE | MAPE | NRMSE | MAPE | NRMSE |
| Airtel | **0.934** | **0.165** | 1.101 | 0.296 | 4.06 | 0.61 | 2.31 | 0.420 | 4.29 | 0.794 | 3.49 | 0.634 | 2.77 | 0.546 |
| Tata Steel | **0.806** | **0.098** | 1.060 | 0.117 | 12.186 | 0.971 | 7.327 | 0.697 | 20.908 | 1.758 | 9.142 | 0.854 | 8.257 | 0.694 |
| TCS | **0.226** | **0.066** | 0.505 | 0.166 | 2.956 | 0.830 | 2.336 | 0.707 | 4.331 | 1.291 | 2.723 | 0.767 | 2.992 | 0.728 |
| SBI | 0.918 | 0.072 | **0.838** | **0.071** | 4.940 | 0.355 | 3.257 | 0.266 | 9.664 | 0.697 | 4.527 | 0.378 | 5.45 | 0.360 |
| ONGC | 0.796 | 0.123 | **0.232** | **0.072** | 1.659 | 0.278 | 0.83 | 0.157 | 3.37 | 0.621 | 1.337 | 0.235 | 2.258 | 0.332 |
| Infosys | **0.567** | **0.097** | 0.865 | 0.181 | 3.665 | 0.63 | 2.95 | 0.531 | 2.536 | 0.408 | 3.87 | 0.721 | 1.912 | 0.402 |
| Sun Pharma | **0.260** | **0.030** | 0.603 | 0.095 | 2.407 | 0.318 | 2.137 | 0.282 | 2.325 | 0.276 | 2.216 | 0.297 | 2.848 | 0.301 |
| Spice Jet | 0.909 | 0.035 | **0.614** | **0.037** | 3.371 | 0.173 | 2.830 | 0.144 | 4.414 | 0.210 | 2.910 | 0.151 | 2.462 | 0.155 |
| Jet Airways | 1.040 | 0.075 | **0.604** | **0.053** | 2.846 | 0.246 | 2.11 | 0.196 | 5.81 | 0.48 | 2.89 | 0.244 | 4.121 | 0.312 |

Table 5. Stock Prediction Results with MRMR (25) feature selection method

| Dataset | GMDH | | GRNN | | RF | | QRRF | | RPART | | SVR | | MLP | |
|---|---|---|---|---|---|---|---|---|---|---|---|---|---|---|
| | MAPE | NRMSE | MAPE | NRMSE | MAPE | NRMSE | MAPE | NRMSE | MAPE | NRMSE | MAPE | NRMSE | MAPE | NRMSE |
| Airtel | **0.355** | **0.069** | 0.614 | 0.149 | 4.45 | 0.761 | 2.83 | 0.511 | 4.34 | 0.737 | 3.68 | 0.666 | 2.202 | 0.462 |
| Mahindra | **1.42** | **0.054** | 6.49 | 0.247 | 14.19 | 0.510 | 12.61 | 0.455 | 17.39 | 0.615 | 16.12 | 0.58 | 12.43 | 0.455 |
| Tata Motors | **0.618** | **0.061** | 0.679 | 0.067 | 6.567 | 0.553 | 4.94 | 0.449 | 5.99 | 0.629 | 6.94 | 0.600 | 4.004 | 0.402 |
| Reliance Industries | **0.435** | **0.066** | 0.698 | 0.115 | 2.516 | 0.351 | 2.129 | 0.299 | 3.346 | 0.426 | 2.139 | 0.312 | 2.327 | 0.291 |
| Tata Steel | **0.654** | **0.064** | 0.786 | 0.075 | 11.311 | 0.927 | 7.130 | 0.687 | 9.78 | 1.031 | 7.96 | 0.801 | 7.713 | 0.706 |
| TCS | **0.279** | **0.078** | 0.522 | 0.167 | 2.860 | 0.785 | 2.950 | 0.695 | 4.111 | 1.164 | 2.813 | 0.763 | 2.26 | 0.621 |
| SBI | 0.918 | 0.072 | **0.838** | **0.071** | 4.865 | 0.356 | 3.207 | 0.254 | 9.664 | 0.697 | 4.527 | 0.378 | 4.85 | 0.331 |
| ONGC | 0.736 | 0.133 | **0.303** | **0.077** | 1.53 | 0.249 | 0.86 | 0.157 | 1.34 | 0.313 | 1.23 | 0.193 | 2.136 | 0.321 |
| Infosys | **0.443** | **0.078** | 0.860 | 0.182 | 3.216 | 0.553 | 2.844 | 0.508 | 3.613 | 0.596 | 3.136 | 0.533 | 2.59 | 0.432 |
| Sun Pharma | **0.320** | **0.035** | 0.517 | 0.074 | 2.357 | 0.296 | 2.026 | 0.265 | 2.868 | 0.327 | 2.644 | 0.350 | 2.529 | 0.253 |
| Spice Jet | 0.838 | 0.034 | **0.294** | **0.014** | 3.120 | 0.166 | 2.005 | 0.124 | 4.105 | 0.212 | 2.774 | 0.137 | 4.444 | 0.179 |
| Jet Airways | 0.975 | 0.068 | **0.404** | **0.047** | 2.98 | 0.263 | 2.41 | 0.226 | 5.04 | 0.422 | 2.669 | 0.212 | 4.497 | 0.331 |

**Table 6. Stock Prediction Results with Chi-square (10) feature selection method**

| Dataset | GMDH | | GRNN | | RF | | QRRF | | RPART | | SVR | | MLP | |
|---|---|---|---|---|---|---|---|---|---|---|---|---|---|---|
| | MAPE | NRMSE | MAPE | NRMSE | MAPE | NRMSE | MAPE | NRMSE | MAPE | NRMSE | MAPE | NRMSE | MAPE | NRMSE |
| Airtel | **1.03** | **0.209** | 1.32 | 0.274 | 4.54 | 0.809 | 2.72 | 0.538 | 6.75 | 1.13 | 3.41 | 0.630 | 3.39 | 0.569 |
| Mahindra | **4.31** | **0.207** | 8.45 | 0.303 | 13.41 | 0.48 | 12.15 | 0.45 | 14.07 | 0.537 | 15.78 | 0.570 | 9.452 | 0.092 |
| Tata Motors | **1.29** | **0.142** | 1.61 | 0.143 | 6.95 | 0.588 | 5.832 | 0.487 | 6.642 | 0.652 | 6.894 | 0.603 | 6.703 | 0.555 |
| Reliance Industries | 1.119 | 0.150 | **1.031** | **0.153** | 2.695 | 0.362 | 2.106 | 0.282 | 2.823 | 0.379 | 2.520 | 0.351 | 2.712 | 0.365 |
| Tata Steel | 1.892 | 0.199 | **1.878** | **0.183** | 12.188 | 0.981 | 9.623 | 0.916 | 16.419 | 1.405 | 10.691 | 0.901 | 9.969 | 0.835 |
| TCS | **0.519** | **0.137** | 0.732 | 0.215 | 2.69 | 0.731 | 1.906 | 0.604 | 2.96 | 0.758 | 2.503 | 0.707 | 1.961 | 0.517 |
| SBI | 1.82 | 0.134 | **1.33** | **0.105** | 5.901 | 0.427 | 4.345 | 0.283 | 8.234 | 0.612 | 6.005 | 0.474 | 5.327 | 0.353 |
| ONGC | 1.406 | 0.214 | **0.351** | **0.080** | 1.565 | 0.254 | 0.685 | 0.135 | 1.88 | 0.311 | 1.327 | 0.219 | 1.954 | 0.301 |
| Infosys | **1.22** | **0.197** | 1.312 | 0.248 | 3.69 | 0.648 | 2.92 | 0.533 | 3.57 | 0.617 | 4.06 | 0.761 | 3.507 | 0.584 |
| Sun Pharma | **0.765** | **0.087** | 1.32 | 0.164 | 2.222 | 0.287 | 1.636 | 0.222 | 2.640 | 0.292 | 2.416 | 0.294 | 2.113 | 0.225 |
| Spice Jet | 2.26 | 0.093 | **1.10** | **0.052** | 2.946 | 0.167 | 2.758 | 0.157 | 4.55 | 0.236 | 3.544 | 0.153 | 3.678 | 0.151 |
| Jet Airways | 2.09 | 0.169 | **1.351** | **0.114** | 3.09 | 0.263 | 2.78 | 0.279 | 6.057 | 0.488 | 3.63 | 0.288 | 4.194 | 0.310 |

**Table 7. Stock Prediction Results with MRMR (10) feature selection method**

| Dataset | GMDH | | GRNN | | RF | | QRRF | | RPART | | SVR | | MLP | |
|---|---|---|---|---|---|---|---|---|---|---|---|---|---|---|
| | MAPE | NRMSE | MAPE | NRMSE | MAPE | NRMSE | MAPE | NRMSE | MAPE | NRMSE | MAPE | NRMSE | MAPE | NRMSE |
| Airtel | **1.86** | **0.347** | 2.44 | 0.424 | 4.505 | 0.78 | 2.47 | 0.446 | 5.41 | 0.96 | 4.50 | 0.82 | 3.09 | 0.618 |
| Mahindra | **2.77** | **0.107** | 8.90 | 0.321 | 13.60 | 0.48 | 12.08 | 0.429 | 13.68 | 0.49 | 15.38 | 0.553 | 8.52 | 0.334 |
| Tata Motors | 1.26 | 0.125 | **1.01** | **0.097** | 6.338 | 0.544 | 4.95 | 0.441 | 4.78 | 0.482 | 6.91 | 0.629 | 4.586 | 0.429 |
| Reliance Industries | 1.582 | 0.208 | **1.063** | **0.152** | 2.510 | 0.351 | 2.112 | 0.298 | 2.842 | 0.449 | 2.272 | 0.350 | 2.407 | 0.348 |
| Tata Steel | **0.933** | **0.097** | 1.120 | 0.1109 | 10.756 | 0.885 | 7.035 | 0.696 | 9.028 | 0.891 | 6.334 | 0.687 | 5.158 | 0.450 |
| TCS | **0.408** | **0.099** | 0.667 | 0.198 | 2.799 | 0.752 | 2.210 | 0.631 | 4.141 | 1.180 | 2.658 | 0.751 | 1.725 | 0.464 |
| SBI | 1.82 | 0.134 | **1.33** | **0.105** | 5.672 | 0.415 | 3.505 | 0.283 | 8.234 | 0.612 | 6.005 | 0.474 | 4.97 | 0.332 |
| ONGC | 1.603 | 0.242 | **0.571** | **0.111** | 1.44 | 0.261 | 0.968 | 0.179 | 2.39 | 0.424 | 1.35 | 0.222 | 2.151 | 0.328 |
| Infosys | 1.209 | 0.197 | **1.047** | **0.203** | 3.124 | 0.518 | 2.746 | 0.479 | 3.13 | 0.491 | 3.012 | 0.491 | 2.49 | 0.419 |
| Sun Pharma | **0.603** | **0.065** | 0.663 | 0.088 | 2.173 | 0.270 | 2.039 | 0.266 | 3.134 | 0.329 | 2.354 | 0.314 | 1.547 | 0.1507 |
| Spice Jet | 1.16 | 0.064 | **0.716** | **0.040** | 3.564 | 0.190 | 2.085 | 0.162 | 5.562 | 0.262 | 2.857 | 0.141 | 2.744 | 0.153 |
| Jet Airways | 1.718 | 0.131 | **0.846** | **0.072** | 3.03 | 0.260 | 3.11 | 0.239 | 6.05 | 0.488 | 3.632 | 0.288 | 4.603 | 0.325 |

**Table 8. Features (LIWC) Selected Through Two Feature Selection Methods**

| Dataset | Feature Selection Method | Features |
|---|---|---|
| Tata Motors | Chi-10 (8) | Analytic, you, quant, female, cogproc, sexual, focuspast, Exclam |
| Tata Motors | MRMR-25 | Focuspresent, quant, health, time, Comma, focuspast, WC, Period, Exclam, you, body, focusfuture, cause, conj, discrep, see, achieve, swear, ipron, male, leisure, posemo, QMark, Apostro, friend |
| Reliance | MRMR-25 | Focusfuture, relig, OtherP, body, space, focuspast, health, drives, motion, friend, Colon, nonflu, Period, Comma, focuspresent, feel, QMark, bio, relativ, Parenth, affiliation, anx, cause, leisure, Dash |
| Tata Steel | Chi-25 | Ipron, sexual, ppron, hear, percept, affiliation, QMark, death, pronoun, feel, shehe, they, we, article, negemo, SemiC, Analytic, male, Apostro, Quote, compare, i, achieve, affect, WPS |
| TCS | Chi-25 | Shehe, relig, WC, OtherP, female, SemiC, Colon, death, differ, sexual, see, i, quant, Dic, AllPunc, male, Comma, achieve, netspeak, interrog, space, certain, QMark, family, adverb |
| Spice Jet | MRMR-10 | Power, adverb, they, Sixltr, Analytic, discrep, AllPunc, relig, Period, SemiC |
| ONGC | Chi-25 | Clout, ipron, pronoun, insight, achieve, feel, Tone, informal, work, health, bio, motion, OtherP, focusfuture, Quote, Apostro, certain, conj, differ, article, drives, prep, cogproc, family, social |
| Jet Airways | MRMR-25 | Certain, Dash, Parenth, motion, we, Sixltr, verb, posemo, home, anger, Analytic, money, risk, Comma, Quote, number, see, article, male, ingest, ppron, relig, WC, family, netspeak |

As seen from Table 9, the absolute value of the DM statistic (Chen et al. 2014) is less than 1.96 in the following cases: Tata Motors (case (iii) and (iv)), Reliance Industries (case (iv)), Tata Steel (case (ii)), TCS (case (ii)) and Spice Jet (case (v)) datasets. It indicates that there is no statistically significant difference between GMDH (case (i)) and GMDH (cases mentioned above) or GRNN (case (i)) and the GRNN (cases referred to above) as the case may be at 5\%level of significance. Therefore, for these datasets, the corresponding cases of feature subset selection methods turned out to be better than the case (i) in terms of MAPE and NRMSE. However, in the rest of the cases in Table 9, the absolute of DM statistic is greater than 1.96 which indicates that case (i) of full features is statistically significantly better than all feature subset selection cases in terms of MAPE and NRMSE at 5% level of significance.

**Table 9. DM Test Values of the Models with LIWC Features**

| Dataset | GMDH (Full features) vs. GMDH ( a/ b/ c/ d) | | | |
|---|---|---|---|---|
| | Chi_25[a] | Chi_10[b] | MRMR_25[c] | MRMR_10[d] |
| Airtel | -2.99 | -2.02 | -2.25 | -2.13 |
| Mahindra | NA | -2.26 | -3.17 | -2.95 |
| Tata Motors | NA | -1.86 | -0.94 | -2.36 |
| Reliance Industries | NA | -3.41 | -1.80 | -3.80 |
| Tata Steel | -1.74 | -2.76 | -2.48 | -2.54 |
| TCS | -1.73 | -3.54 | -2.72 | -3.55 |
| SBI | -2.76 | -2.98 | -2.76 | -2.98 |
| Infosys | -3.21 | -4.25 | -2.92 | -3.63 |
| Sun Pharma | -2.04 | -2.61 | -2.02 | -2.87 |
| Spice Jet | -4.58 | -3.57 | -3.09 | -1.92 |
| | GRNN (Full features) vs. GRNN ( a/ b/ c/ d) | | | |
| ONGC | -0.99 | -2.43 | -4.47 | -2.25 |
| Jet Airways | -2.19 | -2.97 | -1.08 | -2.64 |

NA - Not Applicable

Table 10. Stock Prediction Results with TAALES Full Features

| Dataset | GMDH | | GRNN | | RF | | QRRF | | RPART | | SVR | | MLP | |
|---|---|---|---|---|---|---|---|---|---|---|---|---|---|---|
| | MAPE | NRMSE | MAPE | NRMSE | MAPE | NRMSE | MAPE | NRMSE | MAPE | NRMSE | MAPE | NRMSE | MAPE | NRMSE |
| Airtel | 0.259 | 0.046 | 1.169 | 0.243 | 6.76 | 1.067 | 2.607 | 0.48 | 3.394 | 0.586 | 6.288 | 1.001 | 4.99 | 0.752 |
| Mahindra | 1.782 | 0.073 | 10.34 | 0.349 | 16.91 | 0.592 | 16.431 | 0.578 | 17.57 | 0.618 | 17.19 | 0.605 | 15.85 | 0.579 |
| Tata Motors | 1.372 | 0.141 | 7.17 | 0.613 | 10.25 | 0.827 | 5.41 | 0.495 | 10.05 | 0.829 | 8.97 | 0.748 | 9.092 | 0.729 |
| Reliance Industries | 0.446 | 0.068 | 1.905 | 0.254 | 2.95 | 0.365 | 2.818 | 0.337 | 3.36 | 0.411 | 2.79 | 0.349 | 2.157 | 0.318 |
| Tata Steel | 3.007 | 0.308 | 5.513 | 0.581 | 16.23 | 1.31 | 11.04 | 1.022 | 15.60 | 1.45 | 16.47 | 1.33 | 17.21 | 1.346 |
| TCS | 0.619 | 0.190 | 3.058 | 0.882 | 5.397 | 1.358 | 4.312 | 1.085 | 6.53 | 1.818 | 5.55 | 1.395 | 5.345 | 1.370 |
| SBI | 2.113 | 0.156 | 1.246 | 0.088 | 6.207 | 0.397 | 3.651 | 0.252 | 12.38 | 0.760 | 5.66 | 0.378 | 4.98 | 0.334 |
| ONGC | 0.7112 | 0.113 | 1.055 | 0.1506 | 2.46 | 0.352 | 1.788 | 0.257 | 4.193 | 0.624 | 2.138 | 0.316 | 1.935 | 0.294 |
| Infosys | 0.410 | 0.076 | 4.934 | 0.807 | 4.49 | 0.750 | 3.010 | 0.518 | 5.146 | 0.788 | 4.418 | 0.720 | 4.224 | 0.725 |
| Sun Pharma | 0.867 | 0.084 | 1.960 | 0.208 | 5.49 | 0.494 | 4.46 | 0.403 | 5.43 | 0.497 | 5.198 | 0.472 | 10.95 | 0.998 |
| Spice Jet | 2.090 | 0.098 | 8.023 | 0.303 | 11.623 | 0.40 | 5.135 | 0.244 | 9.756 | 0.395 | 8.252 | 0.317 | 72.33 | 2.37 |
| Jet Airways | 2.014 | 0.178 | 1.842 | 0.153 | 3.95 | 0.311 | 3.79 | 0.275 | 5.714 | 0.464 | 4.295 | 0.328 | 4.507 | 0.339 |

**Table 11. Stock Prediction Results with Ch-25 Features from TAALES Features**

| Dataset | GMDH | | GRNN | | RF | | QRRF | | RPART | | SVR | | MLP | |
|---|---|---|---|---|---|---|---|---|---|---|---|---|---|---|
| | MAPE | NRMSE | MAPE | NRMSE | MAPE | NRMSE | MAPE | NRMSE | MAPE | NRMSE | MAPE | NRMSE | MAPE | NRMSE |
| Airtel | **1.163** | **0.200** | 2.234 | 0.412 | 7.248 | 1.165 | 3.004 | 0.46 | 7.88 | 1.239 | 6.71 | 1.093 | 4.64 | 0.713 |
| Mahindra | **3.53** | **0.151** | 13.37 | 0.487 | 16.99 | 0.594 | 16.183 | 0.547 | 17.63 | 0.609 | 17.67 | 0.614 | 13.85 | 0.509 |
| Tata Motors | **3.74** | **0.367** | 4.45 | 0.437 | 9.102 | 0.773 | 6.109 | 0.550 | 7.99 | 0.777 | 8.15 | 0.744 | 7.757 | 0.656 |
| Reliance Industries | **1.124** | **0.145** | 1.746 | 0.221 | 3.359 | 0.442 | 2.661 | 0.330 | 4.122 | 0.535 | 3.386 | 0.415 | 2.82 | 0.33 |
| Tata Steel | **4.665** | **0.472** | 4.81 | 0.516 | 14.63 | 1.208 | 12.07 | 1.151 | 18.47 | 1.563 | 18.31 | 1.148 | 15.344 | 1.341 |
| TCS | **0.947** | **0.310** | 3.369 | 0.905 | 5.59 | 1.42 | 3.661 | 0.983 | 5.34 | 1.52 | 5.322 | 1.364 | 5.312 | 1.314 |
| SBI | 3.186 | 0.206 | **1.478** | **0.113** | 6.763 | 0.451 | 3.724 | 0.248 | 11.02 | 0.684 | 5.69 | 0.393 | 4.706 | 0.307 |
| ONGC | **1.251** | **0.182** | 1.798 | 0.279 | 3.04 | 0.429 | 1.508 | 0.24 | 2.421 | 0.384 | 2.21 | 0.356 | 2.272 | 0.335 |
| Infosys | **0.697** | **0.123** | 4.696 | 0.727 | 4.582 | 0.749 | 3.676 | 0.613 | 4.599 | 0.734 | 4.308 | 0.678 | 3.872 | 0.658 |
| Sun Pharma | **1.464** | **0.1503** | 2.060 | 0.216 | 5.716 | 0.5008 | 4.608 | 0.433 | 7.35 | 0.619 | 5.060 | 0.457 | 3.46 | 0.355 |
| Spice Jet | **3.017** | **0.174** | 7.572 | 0.284 | 14.109 | 0.469 | 5.809 | 0.265 | 7.697 | 0.346 | 8.245 | 0.317 | 4.341 | 0.206 |
| Jet Airways | **2.0307** | **0.1591** | 2.852 | 0.220 | 5.184 | 0.380 | 3.676 | 0.613 | 5.980 | 0.477 | 5.46 | 0.379 | 4.532 | 0.332 |

**Table 12. Prediction Results with MRMR-25 Features from TAALES Features**

| Dataset | GMDH | | GRNN | | RF | | QRRF | | RPART | | SVR | | MLP | |
|---|---|---|---|---|---|---|---|---|---|---|---|---|---|---|
| | MAPE | NRMSE | MAPE | NRMSE | MAPE | NRMSE | MAPE | NRMSE | MAPE | NRMSE | MAPE | NRMSE | MAPE | NRMSE |
| Airtel | **0.884** | **0.148** | 1.743 | 0.315 | 7.226 | 1.104 | 2.985 | 0.482 | 6.172 | 0.916 | 5.96 | 0.913 | 4.41 | 0.678 |
| Mahindra | **5.037** | **0.202** | 8.98 | 0.304 | 17.07 | 0.596 | 16.18 | 0.569 | 18.14 | 0.634 | 16.52 | 0.580 | 16.51 | 0.573 |
| Tata Motors | **2.93** | **0.278** | 3.626 | 0.298 | 11.341 | 0.89 | 7.084 | 0.600 | 11.11 | 0.966 | 9.538 | 0.789 | 5.835 | 0.524 |
| Reliance Industries | **1.293** | **0.168** | 1.768 | 0.218 | 2.79 | 0.347 | 2.506 | 0.307 | 3.361 | 0.474 | 2.88 | 0.351 | 3.305 | 0.392 |
| Tata Steel | **3.821** | **0.366** | 4.78 | 0.430 | 16.31 | 1.29 | 16.509 | 1.372 | 17.43 | 1.469 | 16.55 | 1.32 | 14.64 | 1.163 |
| TCS | **0.975** | **0.293** | 1.875 | 0.617 | 5.155 | 1.295 | 3.99 | 1.055 | 4.107 | 1.076 | 4.839 | 1.232 | 5.562 | 1.372 |
| SBI | 3.398 | 0.231 | **1.627** | **0.122** | 6.139 | 0.404 | 3.609 | 0.242 | 8.102 | 0.572 | 6.081 | 0.393 | 5.102 | 0.335 |
| ONGC | 1.037 | 0.162 | **0.863** | **0.151** | 3.03 | 0.446 | 1.526 | 0.234 | 4.809 | 0.656 | 2.485 | 0.403 | 2.257 | 0.332 |
| Infosys | **0.845** | **0.143** | 4.858 | 0.749 | 4.901 | 0.791 | 3.884 | 0.643 | 5.373 | 0.837 | 4.013 | 0.654 | 4.372 | 0.681 |
| Sun Pharma | **1.412** | **0.157** | 2.036 | 0.213 | 4.427 | 0.429 | 4.006 | 0.396 | 4.92 | 0.470 | 4.898 | 0.450 | 5.064 | 0.475 |
| Spice Jet | **2.256** | **0.089** | 8.502 | 0.309 | 14.43 | 0.475 | 5.28 | 0.236 | 7.41 | 0.303 | 7.22 | 0.278 | 14.39 | 0.475 |
| Jet Airways | **1.860** | **0.135** | 2.654 | 0.199 | 6.326 | 0.459 | 3.884 | 0.643 | 9.501 | 0.712 | 5.315 | 0.401 | 4.481 | 0.334 |

**Table 13. Stock Prediction Results with Ch-10 Features from TAALES Features**

| Dataset | GMDH | | GRNN | | RF | | QRRF | | RPART | | SVR | | MLP | |
|---|---|---|---|---|---|---|---|---|---|---|---|---|---|---|
| | MAPE | NRMSE | MAPE | NRMSE | MAPE | NRMSE | MAPE | NRMSE | MAPE | NRMSE | MAPE | NRMSE | MAPE | NRMSE |
| Airtel | **1.201** | **0.230** | 2.93 | 0.478 | 7.509 | 1.205 | 5.127 | 0.825 | 8.126 | 1.263 | 7.763 | 1.262 | 4.68 | 0.722 |
| Mahindra | **5.513** | **0.214** | 13.03 | 0.474 | 17.26 | 0.603 | 15.98 | 0.558 | 15.03 | 0.532 | 17.91 | 0.629 | 15.35 | 0.547 |
| Tata Motors | 4.837 | 0.438 | **4.209** | **0.391** | 8.92 | 0.757 | 5.95 | 0.511 | 9.54 | 0.771 | 8.117 | 0.718 | 8.068 | 0.685 |
| Reliance Industries | **1.881** | **0.241** | 2.152 | 0.260 | 3.407 | 0.429 | 2.834 | 0.352 | 3.76 | 0.472 | 3.134 | 0.395 | 3.427 | 0.414 |
| Tata Steel | 7.234 | 0.639 | **5.666** | **0.679** | 15.97 | 1.303 | 14.326 | 1.215 | 13.67 | 1.22 | 20.09 | 1.59 | 13.58 | 1.275 |
| TCS | **1.094** | **0.375** | 3.156 | 0.824 | 5.378 | 1.413 | 4.525 | 1.167 | 6.04 | 1.759 | 5.77 | 1.504 | 5.115 | 1.284 |
| SBI | **2.591** | **0.197** | 6.136 | 1.825 | 6.237 | 0.426 | 3.862 | 0.256 | 10.97 | 0.641 | 5.53 | 0.366 | 5.22 | 0.346 |
| ONGC | 2.115 | 0.306 | **1.160** | **0.200** | 3.37 | 0.477 | 1.61 | 0.246 | 3.93 | 0.545 | 2.79 | 0.436 | 2.278 | 0.334 |
| Infosys | **1.208** | **0.211** | 4.587 | 0.712 | 4.751 | 0.772 | 4.283 | 0.700 | 6.245 | 1.084 | 4.682 | 0.726 | 4.413 | 0.681 |
| Sun Pharma | **1.464** | **0.150** | 2.93 | 0.303 | 5.69 | 0.505 | 5.371 | 0.480 | 7.30 | 0.645 | 5.26 | 0.477 | 5.053 | 0.477 |
| Spice Jet | **3.101** | **0.161** | 6.381 | 0.262 | 13.23 | 0.444 | 7.973 | 0.295 | 7.582 | 0.342 | 8.74 | 0.314 | 4.635 | 0.230 |
| Jet Airways | 2.605 | 0.204 | **2.224** | **0.169** | 4.92 | 0.349 | 4.283 | 0.700 | 7.281 | 0.564 | 5.73 | 0.405 | 4.541 | 0.334 |

## Table 14. Prediction Results with MRMR-10 Features from TAALES Features

| Dataset | GMDH | | GRNN | | RF | | QRRF | | RPART | | SVR | | MLP | |
|---|---|---|---|---|---|---|---|---|---|---|---|---|---|---|
| | MAPE | NRMSE | MAPE | NRMSE | MAPE | NRMSE | MAPE | NRMSE | MAPE | NRMSE | MAPE | NRMSE | MAPE | NRMSE |
| Airtel | **2.51** | **0.487** | 3.06 | 0.560 | 7.013 | 1.098 | 2.818 | 0.569 | 8.045 | 1.255 | 7.018 | 1.125 | 4.344 | 0.667 |
| Mahindra | **7.32** | **0.281** | 10.23 | 0.351 | 18.15 | 0.629 | 16.21 | 0.572 | 18.88 | 0.664 | 17.33 | 0.609 | 16.75 | 0.585 |
| Tata Motors | **4.462** | **0.391** | 5.24 | 0.444 | 9.691 | 0.795 | 6.668 | 0.571 | 11.38 | 1.01 | 9.57 | 0.807 | 7.32 | 0.647 |
| Reliance Industries | 1.857 | 0.246 | **1.788** | **0.235** | 3.091 | 0.375 | 2.557 | 0.305 | 3.99 | 0.520 | 2.97 | 0.352 | 2.59 | 0.323 |
| Tata Steel | **6.448** | **0.592** | 6.818 | 0.655 | 15.92 | 1.288 | 12.62 | 1.123 | 15.83 | 1.33 | 17.06 | 1.37 | 11.26 | 0.976 |
| TCS | **1.682** | **0.541** | 3.107 | 0.945 | 5.483 | 1.39 | 3.97 | 1.004 | 7.062 | 1.821 | 5.591 | 1.41 | 5.461 | 1.374 |
| SBI | 4.758 | 0.318 | **3.053** | **0.215** | 6.91 | 0.474 | 3.686 | 0.253 | 9.77 | 0.667 | 6.61 | 0.436 | 5.563 | 0.365 |
| ONGC | 1.921 | 0.326 | **1.352** | **0.205** | 2.62 | 0.391 | 1.706 | 0.262 | 3.93 | 0.56 | 2.45 | 0.370 | 2.269 | 0.332 |
| Infosys | **1.0301** | **0.180** | 4.868 | 0.748 | 5.184 | 0.802 | 3.911 | 0.617 | 7.014 | 1.098 | 4.648 | 0.742 | 4.115 | 0.675 |
| Sun Pharma | 2.99 | 0.312 | **2.72** | **0.267** | 4.781 | 0.453 | 4.132 | 0.395 | 5.24 | 0.495 | 5.431 | 0.504 | 5.60 | 0.531 |
| Spice Jet | **2.734** | **0.148** | 8.509 | 0.309 | 11.978 | 0.404 | 5.16 | 0.241 | 6.784 | 0.295 | 6.91 | 0.270 | 12.74 | 0.431 |
| Jet Airways | **2.146** | **0.203** | 3.776 | 0.280 | 6.85 | 0.493 | 3.911 | 0.617 | 8.92 | 0.628 | 5.67 | 0.429 | 4.352 | 0.322 |

Table 10 presents the MAPE and NRMSE values yielded on TAALES features by various prediction models without feature selection. It can be observed that except on one dataset (JetAirways), GMDH outperformed the other techniques on all the datasets in terms of MAPE and NRMSE. For the JetAirways dataset, GRNN performed better than GMDH. Table 11 presents the results obtained by employing various models on the top-25 features obtained by Chi-square feature selection method. From Table 11, we can observe that except for the SBI dataset, GMDH yielded the best predictions for all datasets. For this dataset, GRNN outperformed all other techniques. The results obtained with top-25 features selected by MRMR method are presented in Table 12. In this combination, GMDH performed the best in terms of MAPE and NRMSE on all datasets except SBI, and ONGC datasets, for which, GRNN performed the best. Table 13 summarizes the results yielded by the models on the top-10 features selected by Chi-square method. In this table, it can be observed that the GMDH outperformed all other techniques on the datasets of Airtel, Mahindra, Reliance Industries, TCS, SBI, Infosys, Sun Pharma, Spice Jet. Whereas, GRNN yielded the best predictions on Tata Motors, Tata Steel, ONGC and Jet Airways in terms of both MAPE and NRMSE values. Table 14 presents the results obtained with top-10 features selected by MRMR method. In this table, it can be observed that GRNN outperformed other models in terms of both MAPE and NRMSE on four companies' stocks (Reliance Industries, SBI, ONGC, and Sun Pharma) and GMDH performed the best on the remaining eight datasets.

Table 15. DM Test Values of the Models with TAALES Features

| Dataset | GMDH (Full features) vs. GMDH ( a/ b/ c/ d) | | | |
|---|---|---|---|---|
| | Chi_25[a] | Chi_10[b] | MRMR_25[c] | MRMR_10[d] |
| Airtel | -3.5448 | -1.9618 | -2.9054 | -2.2615 |
| Mahindra | -2.3672 | -2.9913 | -2.8336 | -3.583 |
| Tata Motors | -2.3954 | -2.6174 | -3.0627 | -3.662 |
| Reliance Industries | -3.529 | -3.8573 | -4.222 | -3.9516 |
| Tata Steel | -2.0483 | -4.6712 | **-1.0015** | -3.5876 |
| TCS | **-1.8943** | **-1.8918** | -2.22 | -2.4727 |
| SBI | **-1.6882** | **-0.9737** | -2.1516 | -3.2354 |
| ONGC | -2.4633 | -3.9292 | **-1.6057** | -2.8267 |
| Infosys | **-1.6201** | -3.1324 | -2.6704 | -3.0838 |
| Sun Pharma | -3.2006 | -3.0988 | -2.0877 | -3.5302 |
| Spice Jet | **-1.9323** | -2.2338 | **0.40826** | -2.1279 |
| | GRNN (Full features) vs. GRNN ( a/ b/ c/ d) | | | |
| Jet Airways | **-1.5835** | **-0.6087** | -2.188 | -3.2475 |

Using the features extracted using TAALES as the feature subset, a statistical significance test called Diebold-Mariano Test (DM) was conducted between case (i) (all features) of GMMDH and all other cases of GMDH in a pair wise manner for all datasets except Jet Airways. For this dataset, we performed DM test between case (i) of GRNN and all other cases of GRNN in a pair wise manner. We chose GMDH and GRNN because of their superior performance over other models in terms of MAPE and NRMSE as seen in Tables 10 through 14. The DM test values are reported in Table 15. As seen in Table 15, the absolute value of the DM statistic (Chen et al., 2014) is less than 1.96 in the following cases. Tata Steel (case (iv)), TCS (case (ii) and (iii)), SBI (case (ii) and (iii)), ONGC (case (iv)), Infosys (case(ii)), Spice Jet (case (ii) and (iv)), and Jet Airways (case (ii) and (iii)). It indicates that there is no statistically significant difference between GMDH (case (i)) and GMDH (cases mentioned above) or GRNN (case (i)) and the GRNN (cases referred to above) as the case may beat 5\% level of significance. Therefore, in these datasets, the corresponding cases of feature subset selection methods turned out to better than the case (i) in terms of MAPE and NRMSE. However, in the rest of the cases in Table 15, the absolute of DM statistic is greater than 1.96 which indicates that case (i) of full features is statistically significantly better than all feature subset selection cases in terms of MAPE and NRMSE at 5% level of significance.

Similarly, we also conducted the Diebold-Mariano test (DM) between the LIWC and TAALES models i.e. Between case (i) (all features) of GMDH (LIWC) and case (i) (all features) of GMDH (TAALES) in a pair wise manner for all datasets.

**Table 16. DM Test Values of LIWC vs. TAALES Feature Models**

|  | Full Features | |
|---|---|---|
| **Dataset** | **GMDH (LIWC) vs. GMDH (TAALES)** | **GRNN (LIWC) vs. GRNN (TAALES)** |
| Airtel | -2.3504 | 0.74582 |
| Mahindra | -3.1204 | 0.54479 |
| Tata Motors | -2.1748 | -5.2688 |
| Reliance Industries | -1.4084 | -3.183 |
| Tata Steel | -3.385 | -2.5014 |
| TCS | -2.9328 | -3.591 |
| SBI | -2.7933 | -2.5652 |
| ONGC | -1.6129 | -2.4851 |
| Infosys | -2.4362 | -4.1836 |
| Sun Pharma | -4.105 | -2.863 |
| Spice Jet | -2.2874 | -5.3706 |
| Jet Airways | -2.0484 | -3.3333 |

Similarly, case (i) of GRNN (LIWC) vs. case (i) of GRNN (TAALES). We chose GMDH and GRNN because of their superior performance over other models in terms of MAPE and NRMSE as seen in Tables 3 and 10. The DM Test values are reported in Table 16.

As seen in Table 16, the absolute value of the DM statistic is less than 1.96 in the following cases: Airtel (GRNN), Mahindra (GRNN), Reliance Industries (GMDH), ONGC (GMDH). It indicates that there is no statistically significant difference between GMDH (LIWC) and GMDH (TAALES) with case (i) or GRNN (LIWC) and GRNN (TAALES) with case (i) at 5% level of significance. Therefore, for these datasets, the corresponding models of both, TAALES and LIWC turned out to be equally good in case (i), in terms of MAPE and NRMSE. However, in the rest of the cases in Table 16, the absolute of DM statistic is greater than 1.96 which indicates that case (i) of full features with GMDH or GRNN is statistically significantly better than all feature subset selection cases in terms of MAPE and NRMSE at 5% level of significance.

Thus, the two different feature selection methods adopted here did not perform uniformly well on all datasets because they are filter based approaches and are not as powerful as the wrapper based ones. In this context, one can employ the new elitist quantum inspired differential evolution based wrapper developed by Srikrishna et al. (2015) to see if any significant improvement in prediction accuracy can be obtained. The reason for this suggestion is that, it not only depends on the impressive search capabilities of Differential Evolution but, also the powerful quantum computing principles.

## 8. Conclusions and Future directions

In this paper, a novel stock market prediction model based on the psycholinguistic features extracted from selected, stock (company) related, news articles, is proposed. Various prediction models viz., RF, QRRF, GMDH, SVR, CART, MLP and GRNN were employed for regression. Experiments were conducted on stock prices of 12 companies listed on BSE. Due to non-availability of news articles of some days, for a particular stock, mean-distance based data imputation was employed. In our experiments, it was found that statistically, GMDH yielded the best performance followed by GRNN in terms of MAPE and NRMSE using the DM test. LIWC features models are performing better as compared to TAALES features models. Going further, technical indicators can also be included as predictor variables along with the psycholinguistic and lexical features to get higher accuracies. It is important to note that in the current research, we employed filter-based feature subset selection methods. However, wrapper-based feature subset selection methods, which are designed to take inter-variable interaction effects into consideration may prove to be more potent and are worth exploring. Further, ensembling the predictions yielded by some well performing intelligent techniques is also a future research direction. Finally, psycholinguistic features coupled with evolutionary computation based stock prediction models (Jayakrishna and Ravi, 2016) is another direction worth exploring.